\documentstyle[tighten,12pt,aps]{revtex}
\begin{document}
\draft
\title{X-ray photoemission spectroscopy of La$_{0.67}$Ca$_{0.33}$MnO$_{3}$ films}
\author{P. R. Broussard, S. B. Qadri and V. C. Cestone}
\address{Naval Research Lab, Washington, DC 20375}
\bigskip 
\maketitle
\begin{abstract}
We have performed x-ray photoemission spectroscopy (XPS) on thin films of 
(001) and (200) oriented La$_{0.67}$Ca$_{0.33}$MnO$_{3}$ grown on (100) 
and (110) SrTiO$_{3}$ 
substrates by off-axis sputtering.  The films were examined by XPS without 
exposing them to air.  We have compared the core levels and the valence 
spectra between the two different orientations, as well as after the 
effects of air exposure and annealing in UHV.  We find that the surfaces 
are very stable against exposure to air.  Comparing the measured intensity 
ratios to a model for the uniform termination of the film shows the 
terminating layer to be MnO$_{2}$ for both the (001) and (200) oriented 
La$_{0.67}$Ca$_{0.33}$MnO$_{3}$ films.
\end{abstract}
\bigskip


\narrowtext

\section{Introduction}¥
The colossal magnetoresistance (CMR) behavior seen in thin films of the 
lanthanum based manganese oxides\cite{Jin} has resulted in a great deal of 
interest in thin film growth of the materials.  X-ray photoemission studies 
have been carried out on these systems to understand the change in 
electronic structure as a function of 
doping.\cite{Taguchi,Chainani,Saitoh} Valence-band 
structure has also been studied for a La$_{0.65}$Ca$_{0.35}$MnO$_{3}$ film using 
angle-resolved photoemission;\cite{Zhang} however, there has been no x-ray 
photoemission spectroscopy (XPS) studies of the core states for thin films 
of these materials.  In this paper, we present {\it in situ} and {\it 
ex situ} XPS 
measurements for films of La$_{0.67}$Ca$_{0.33}$MnO$_{3}$ (LCMO) grown 
on SrTiO$_{3}$ 
substrates.  We have examined how the core lines are affected by exposure 
to air and subsequent annealing in UHV, and we have compared our measured 
intensities to a quantitative model for the terminating surface of the 
films.  We have also compared the differences between LCMO films grown on 
different orientations of SrTiO$_{3}$.

\section{Sample preparation and characterization}
Our samples were grown by off-axis sputtering using a composite target of 
LCMO material mounted in a copper cup.  The substrates were (100) and (110) 
oriented strontium titanate (SrTiO$_{3}$), silver-pasted onto a stainless steel 
substrate holder that was radiatively heated from behind by quartz lamps.  
Although there was no direct measurement of the holder temperature for the 
runs used in this study, previous runs (under nominally the same 
conditions) using a thermocouple clamped onto the front surface of the 
holder indicated a temperature of 670 C.  The films were rf-sputtered at 
80 W in a sputter gas composed of 80\%  Ar and 20\%  O$_{2}$ (as measured by flow 
meters) and at a total pressure of 13.3 Pa.  These conditions gave 
deposition rates of $\approx$ 17 nm/hr, with film thicknesses being typically 50 
nm.  After deposition, the samples were cooled in 13.3 kPa of oxygen.  
After the samples had cooled to below 100 C, the chamber was pumped out to 
below 10$^{-4}$ Pa, and the samples were moved by a manipulator arm into an 
adjacent chamber that has the XPS analytical equipment.  After the sample 
was placed into the XPS stage, the chambers were isolated, and the XPS 
chamber pressure evacuated below 3 x 10$^{-7}$ Pa.  For characterization after 
exposure to air, the sample surfaces were cleaned by UHV anneals to 600 C.  
Such anneals are known to have a detrimental effect on the oxygen content 
in other perovskite materials (such as the high T$_{c}$ superconductor 
YBa$_{2}$¥Cu$_{3}$¥O$_{7-\delta}$).  For the case of LCMO, we have seen in 
anneals on {\it in situ} films 
(not shown here) and in our {\it ex situ} film results (shown later) that such 
anneals do not cause any significant reduction in the measured oxygen 
levels.

The XPS spectra were taken at room temperature with a Vacuum Generators 
CLAM 100 analyzer using Mg K$\alpha$ radiation (12 kV, 20 mA).  The emission angle 
of the photoelectrons was 15$^{\circ}¥$ to the surface normal, unless otherwise 
stated.  The pass energy was set to 50 eV, giving a full width at half 
maximum (FWHM) of 2.0 eV for a Pt 4f$_{7/2}$ line.  For the transmission 
function we use what is commonly seen in VG ESCALab detectors 
($T(E_{k}¥) \propto
1/\sqrt{E_{k}¥}$), which are also 150$^{\circ}¥$ sector angle hemispherical analyzers.

The collected XPS peaks were fit by multiple Gaussian curves after 
subtracting away an S-shaped background\cite{Shirley} to obtain the peak positions 
and peak areas.  The energy scale for the system was calibrated using an 
ion-milled Pt foil sample, and has an absolute accuracy of $\pm$0.5 eV and a 
relative energy accuracy of $\pm$0.1 eV for all samples.

In addition to the XPS studies, the samples were characterized {\it ex situ} by 
standard $\theta-2\theta$ x-ray diffraction scans, electrical resistivity measurements 
(using the van der Pauw method\cite{VdP}) for zero applied field, and 
magnetization measurements at low fields using a Quantum Design SQUID 
Magnetometer.

\section{Results and Discussion}¥
\subsection{Film properties}
On (100) and (110) SrTiO$_{3}$ we get smooth, well-oriented LCMO films.  In 
Figure \ref{XRD} we show the x-ray diffraction patterns along the film growth 
direction for LCMO films on (100) and (110) SrTiO$_{3}$.  We find (001) LCMO on 
(100) SrTiO$_{3}$, with a c-axis lattice constant of 7.60\AA .  For the LCMO film 
on (110) SrTiO$_{3}$, we had to use x-ray diffraction in the plane of the film 
since it can be difficult to distinguish between the (200) and (112) LCMO 
reflections.  What we find for this film is that it is oriented as (200) 
LCMO, with an a-axis lattice constant of 5.44\AA .  Furthermore, we find that 
the LCMO is oriented in the film plane with (001) LCMO $\parallel$ (001) 
SrTiO$_{3}$ and 
(010) LCMO $\parallel$ (1$\bar{1}$0) SrTiO$_{3}$, with b and c-axis lattice constants of 
5.52 and 7.77\AA , respectively.  
SEM micrographs (not shown) show no evidence of second 
phases or particulates.  In Figure \ref{rho} we present the resistivity data in 
zero applied magnetic field and magnetization data (at 50 Oe) for the LCMO 
films on the two substrates.  We see a maximum in resistivity near 200 K 
and a magnetic transition that occurs slightly above the resistivity peak 
in temperature.

\subsection{XPS characterization}
\subsubsection{{\it In situ} vs. {\it ex situ} studies}¥
Photoemission data for all the samples was collected around the Mn 2p 
doublet, the La 4d and 3d doublet, the O 1s, Ca 2p, C 1s, Mn 3p lines, and 
the valence region.  In the following figures of XPS spectra, the core 
level spectra are shown along with their deconvolution into different 
Gaussian contributions.  In Figures \ref{XPS1} and \ref{XPS2}, we show scans around the 
location of the above core lines and valence region for a LCMO film on 
(100) SrTiO$_{3}$ after insertion into the spectrometer (A), after exposure to 
room air for approximately 5 minutes (B), and after annealing to 600 C for 
1 hour in UHV (C).  The curves are offset for clarity.

For the {\it in situ} measurement (A) we see no evidence of carbon on the surface 
of the film, therefore no cleaning procedures were needed for the {\it in situ} 
studies.  The {\it in situ} surfaces have not undergone any ion milling or 
thermal cycling, and are thus representative of what would be used in 
further film processing.  The O 1s line has a main peak at 529.1 eV with a 
broad satellite of $\approx$ 71 \% intensity located at 529.5 eV.  The Mn 
2p$_{3/2}$ line 
is also composed of a low binding energy component at 641.3 eV and a high 
binding energy component with $\approx$ 90\% intensity at 643.3 eV.  The spin orbit 
splitting of the Mn 2p lines is 11.7 eV, similar to that seen in 
MnO$_{2}$,\cite{Perkin} 
but our binding energy is lower.  The La 3d$_{5/2}$ line is well fit by a double 
Gaussian, with peaks at 833.7 eV and 837.4 eV, and with the high binding 
energy peak having 115 \% of the intensity of the low peak.  The La 3d 
series is very similar to that of La$_{2}$O$_{3}$, with our spin orbit splitting 
being 16.7 eV.\cite{Perkin} However, as in the case of the Mn 2p lines, our binding 
energy is lower than that seen in La$_{2}$O$_{3}$.  The Mn 3p line is similar to the 
Mn 2p$_{3/2}$ line, with a high binding energy component of 119\% intensity of 
the low binding energy component, and with peak positions of 50.1 and 48.3 
eV respectively.  The La 4d doublet is composed of two spin-orbit doublets 
with binding energies of the 4d$_{5/2}$ line at 101.9 and 104.1 eV, with the 
higher binding energy component having 122\% of the intensity of the lower 
component.  The Ca 2p doublet, in contrast to the other lines, is well fit 
by a standard p-like doublet, with no need for satellite terms.  The Ca 
2p$_{3/2}$ line is located at 346.3 eV.  Finally, the valence region shows 
contributions from the La 5p, O 2s, Ca 3p, and La 5s lines, as well as Mn-O 
bonding below 10 eV.  We find overall good agreement when we compare the 
binding energies of our core levels to that seen by Taguchi et al.  for 
bulk La$_{1-x}$Ca$_{x}$MnO$_{3}$.\cite{Taguchi}

For the case of a 5 minute exposure to room air (B), we see the appearance 
of a carbon line at 285 eV, indicating the presence of hydrocarbons on the 
surface, along with a reduction in intensity of the remaining lines.  
Comparing the lines to those for the {\it in situ} case (A), we do not see a 
great deal of modification in the lineshapes.  A numerical comparison shows 
the greatest modification to be in the O 1s line, where there is an 
increase of 30\% in the relative intensity of the high binding energy 
component compared to the low binding energy peak.  For the valence 
structure, a peak near 9-10 eV is usually associated with 
contamination.\cite{Saitoh}
We do not observe a change in this region after exposing the sample to air; 
however it may be that this peak does not have a significant cross section 
for Mg K$\alpha$ radiation.  After annealing the film to 600 C in UHV (C), we see 
a reduction in the carbon level, and again, no major modification in the 
line shapes.  What we find most interesting about this series is that the 
surface of the LCMO film seems very robust and does not undergo a rapid 
reaction under exposure to room air.

\subsubsection{Surface layer issues}¥
In XPS studies on the high temperature superconductors,\cite{Ziegler} it was customary 
to assign the high binding energy component seen in the core lines with a 
surface layer having a different chemical environment than the bulk.  By 
making XPS measurements at different angles to the surface normal, it is 
possible to distinguish bulk and surface contributions to a photoemission 
line.  We have compared scans of the La 3d$_{5/2}$, Mn 2p$_{3/2}$, O 1s, 
and La 4d 
lines at two different angles, $\theta$, where $\theta$ is the angle between 
the sample 
normal and the plane defined by the x-ray source and detector.  Due to 
limitations of the sample holder, the maximum angle we were able to use was 
45$^{\circ}$.  In Figure \ref{O1sXPS} we show a comparison for the O 1s line for 
our {\it in situ} 
film (A).  What we find here is true for the other lines: there is very 
little change in the lineshape as one goes to different angles.  If either 
component were a surface feature, we would have expected at least a 
60\%  
change in relative intensity ratio of the two components.  Even for the 
case of the LCMO film exposed to air for 5 minutes (B), where we saw an 
increase in the relative intensity of this high binding energy peak for the 
O 1s line and would interpret it as a possible surface component, we find 
that there is very little change between the 0$^{\circ}$ and 45$^{\circ}$ spectra.  This 
implies that the high binding energy components are not due to surface 
layers, and instead are representative of the bulk of the LCMO material.

\subsubsection{Effects of different orientations and termination study}¥
We also performed {\it in situ} XPS measurements for the LCMO film grown on (110) 
SrTiO$_{3}$.  What we observe first for this film is that the core lines are 
nearly identical to our LCMO film on (100) SrTiO$_{3}$, with only the relative 
intensity ratios varying.  So rather than present the curves for this 
sample, we instead look at the variation in intensity ratios, which are 
presented in Table \ref{Table1}.  To study the variation quantitatively, we use the 
model of Frank et al.\cite{Frank} (applied here to the case of LCMO) which models 
the XPS intensities as being due to a perfect termination of the film 
surface and takes into account the contribution from each layer.  For the 
crystal structure of LCMO we used data from LaMnO$_{3}$ based 
structures\cite{Gilleo,Elemans} 
and our values of the lattice parameters.  For the films on (100) SrTiO$_{3}$ we 
use a sequence of layers along the (001) direction, which gives a 
.../O/Mn/O/La$_{0.67}$Ca$_{0.33}$O/...  sequence.  However, in our calculation, 
the 
difference between this more exact sequence compared to the approximate 
sequence .../MnO$_{2}$/La$_{0.67}$Ca$_{0.33}$O/...  is not significant, so we 
have used 
the latter sequence.  We then have two possible configurations, S$_{(001)}$1 
(MnO$_{2}$/La$_{0.67}$Ca$_{0.33}$O/...) 
and S$_{(001)}$2 (La$_{0.67}$Ca$_{0.33}$O/MnO$_{2}$/...).  For the 
film on (110) SrTiO$_{3}$ we also have only two possible configurations for 
(100) oriented LCMO, S$_{(100)}$1 (La$_{0.67}$Ca$_{0.33}$MnO/O$_{2}$/...) 
and S$_{(100)}$2 
(O$_{2}$/La$_{0.67}$Ca$_{0.33}$MnO/...).  The values for photoelectron cross sections and 
inelastic mean free path are taken from the usual 
sources.\cite{Scofield,Tanuma} The 
calculated values of relative intensities for a specified termination layer 
were then compared to the experimental values and the termination layer 
varied to minimize $\chi^{2}$, where
\begin{equation}
\chi^{2}=\frac{1}{N} \sum_{i} \frac{(R_{i,m}-R_{i,p})^{2}}{R_{i,m}} ,
\end{equation}
with N the number of ratios compared, R$_{i,m}$ the measured intensity ratio and 
R$_{i,p}$ the predicted intensity ratio for peak ÒiÓ.  The results of this 
analysis are shown in Tables \ref{Table2} and \ref{Table3} for the LCMO films 
on different 
orientations of SrTiO$_{3}$, where we also compare the intensity ratios to what 
would be expected for a random matrix.

For the films on (100) SrTiO$_{3}$, we see a clear preference for the films 
terminating in a MnO$_{2}$ configuration.  The scans taken at 45$^{\circ}$ also show a 
significant enhancement of the Mn and O signal levels, which would imply a 
MnO$_{2}$ surface layer.  However, what we see for all the samples on (100) 
SrTiO$_{3}$ is a significantly higher O 1s/La 3d$_{5/2}$ ratio than that predicted, 
even for case (C) for which the specimen was annealed in UHV to 600 C.  If 
we assume that the top layer has an enhanced oxygen content, we would 
require an increase in the oxygen level of 150\% to get agreement with our 
measured values.  Clearly this is not reasonable.  Requiring a higher 
oxygen content throughout the film does not give a more reasonable picture, 
especially in consideration of the fact that our films are oxygen-deficient 
as shown by annealing them in oxygen.  The reason for the substantially 
increased O 1s signal level is not understood.

For the film on (110) SrTiO$_{3}$ we find that the values of 
$\chi^{2}$ for the S$_{(100)}$ 
models are much poorer than those for the S$_{(001)}$ models on (100) SrTiO$_{3}$, 
and that the random model gives a better agreement.  If we compare to the 
models for (001) oriented LCMO, however, we find that the best fit is with 
the S$_{(001)}$1 termination, or the MnO$_{2}$ layer.  This would not be expected 
given the observed orientation of the film, and would seem to imply that 
the MnO$_{2}$ layer is an energetically favorable free surface.  What maybe 
happening is that the Mn atoms tend to form a MnO$_{2}$ terminating layer, 
causing a distortion in the layers near the top surface, which might 
explain the low value of $\chi^{2}$ found for these fits.  Attempts to model 
a MnO$_{2}$ 
layer artificially placed onto a (100) oriented LCMO film also gave good 
results, but not better than that seen for the S$_{(001)}$1 termination.

In Figure \ref{valence} we present detailed scans of the valence region for our {\it in situ} 
films on (100) SrTiO$_{3}$ and (110) SrTiO$_{3}$.  For the film on (100) SrTiO$_{3}$ we 
see two dominant contributions near 4 and 6.5 eV.  The data are similar to 
that seen by Chainani et al.\cite{Chainani} for 
La$_{1-x}$Sr$_{x}$MnO$_{3}$, but are less similar to 
the data on a La$_{0.65}$Ca$_{0.35}$MnO$_{3}$ film taken at 50 eV photon 
energy by Zhang 
et al.\cite{Zhang} In that work the dominant peak was seen at 5.8 eV compared 
to our 
6.5 eV.  Zhang et al. also see a high binding energy shoulder at 7.8 eV.  
For the film on (110) SrTiO$_{3}$ we see that there is a loss of intensity at 
the peak positions seen for the film on (100) SrTiO$_{3}$.  It might be 
initially assumed that the change in the spectrum is due to the different 
crystal orientation.  However, in light of the results from the termination 
study it is more likely that what we are seeing here is the result of the 
distortion present in the top surface as it orders in a MnO$_{2}$ structure.  If 
the Mn atoms come from deeper layers, the local distortion in the Mn-O 
bonds may result in a smearing of the local density of states, resulting in 
a valence curve as seen on (110) SrTiO$_{3}$.

\section{Conclusions}
We find from quantitative analysis of XPS data that {\it in situ} films of LCMO 
on both (100) and (110) SrTiO$_{3}$ terminate with MnO$_{2}$ layers, even though the 
film on (110) SrTiO$_{3}$ has an orientation clearly different than the film on 
(100) SrTiO$_{3}$.  We do not see any significant difference in the binding 
energies between the two different films, but only a variation in the 
relative intensity ratios for the core levels.  There is a change in the 
valence structure, which we feel is due to a possible distortion in the top 
surface of the LCMO film grown on (110) SrTiO$_{3}$.  Exposure to air for 5 
minutes and subsequent annealing at 600 C in UHV does not significantly 
modify the shape of the core levels or the valence structure.  Comparing 
scans at 0$^{\circ}$ and 45$^{\circ}$ to the surface normal shows that the core level peaks 
are not composed of a surface and bulk component, but are representative of 
the bulk.

\begin{figure}
\caption{X-ray diffraction scans along the film normal for LCMO films grown 
on (a) (100) SrTiO$_{3}$ and (b) (110) SrTiO$_{3}$.  The reflections due to the 
substrate are labeled by ÒsÓ.}
\label{XRD}
\end{figure}

\begin{figure}
\caption{(a) Electrical resistivity and (b) relative magnetization vs.  
temperature for LCMO films grown on (100) SrTiO$_{3}$ and (110) SrTiO$_{3}$.}
\label{rho}
\end{figure}

\begin{figure}
\caption{XPS spectra and peak fit analysis of (a) C 1s and La 4s, (b) O 1s, 
(c) Mn 2p$_{3/2}$, and (d) La 3d$_{5/2}$ lines.  Spectra are shown for (A) 
an {\it in situ} 
film, (B) after exposure to air for 5 minutes, and (C) after annealing to 
600 C in UHV.  All are for LCMO on (100) SrTiO$_{3}$.}
\label{XPS1}
\end{figure}

\begin{figure}
\caption{Same as Figure 3 but for (a) Mn 3p, (b) La 4d, (c) Ca 2p, and (d) 
the valence region.}
\label{XPS2}
\end{figure}

\begin{figure}
\caption{XPS spectra around the O 1s line for the {\it in situ} film on (100) 
SrTiO$_{3}$ (A) taken at 0$^{\circ}$ and 45$^{\circ}$ relative to the 
surface normal.}
\label{O1sXPS}
\end{figure}

\begin{figure}
\caption{XPS spectra in the valence region for {\it in situ} LCMO films grown on 
(100) and (110) SrTiO$_{3}$.  The curves are offset for clarity.}
\label{valence}
\end{figure}

\begin{table}
\caption{Comparison of relative XPS intensities for LCMO films}
\label{Table1}
\begin{tabular}{cccccc}
Sample & $\frac{Mn 2p_{3/2}}{La3d_{5/2}}$ & $\frac{O 
1s}{La3d_{5/2}}$ & $\frac{Ca 2p}{La3d_{5/2}}$ & $\frac{La 
4d}{La3d_{5/2}}$ & $\frac{Mn 3p}{La3d_{5/2}}$ \\
\hline
(A) {\it In situ} on (100) SrTiO$_{3}$ & 0.51 & 0.45 & 0.087 & 0.26 & 
0.050 \\
(B) {\it Ex situ} on (100) SrTiO$_{3}$ & 0.49 & 0.49 & 0.11 & 0.28 & 
0.054 \\
(C) 600 C UHV (100) SrTiO$_{3}$ & 0.49 & 0.46 & 0.098 & 0.27 & 
0.051 \\
{\it In situ} on (110) SrTiO$_{3}$ & 0.54 & 0.56 & 0.13 & 0.28 & 
0.063 \\
\end{tabular}
\end{table}

\begin{table}
\caption{Calculated Intensity Ratios for (001) oriented LCMO}
\label{Table2}
\begin{tabular}{ccccccccc}
Model & $\frac{Mn 2p_{3/2}}{La3d_{5/2}}$ & $\frac{O 
1s}{La3d_{5/2}}$ & $\frac{Ca 2p}{La3d_{5/2}}$ & $\frac{La 
4d}{La3d_{5/2}}$ & $\frac{Mn 3p}{La3d_{5/2}}$ & $\chi^{2}$(A) & 
$\chi^{2}$(B) & $\chi^{2}$(C) \\
\hline
S$_{(001)}$1 & 0.51 & 0.35 & 0.091 & 0.23 & 
0.079 &  0.0090 & 0.014 & 0.0109 \\
S$_{(001)}$2 & 0.35 & 0.27 & 0.083 & 0.20 & 
0.058 &  0.028 & 0.035 & 0.028 \\
Random & 0.44 & 0.32 & 0.094 & 0.24 & 
0.077 &  0.012 & 0.018 & 0.014 \\
\end{tabular}
\end{table}

\begin{table}
\caption{Calculated Intensity Ratios for (200) oriented LCMO}
\label{Table3}
\begin{tabular}{ccccccc}
Model & $\frac{Mn 2p_{3/2}}{La3d_{5/2}}$ & $\frac{O 
1s}{La3d_{5/2}}$ & $\frac{Ca 2p}{La3d_{5/2}}$ & $\frac{La 
4d}{La3d_{5/2}}$ & $\frac{Mn 3p}{La3d_{5/2}}$ & $\chi^{2}$ \\
\hline
S$_{(100)}$1 & 0.41 & 0.28 & 0.081 & 0.20 & 
0.062 &  0.044 \\
S$_{(100)}$2 & 0.43 & 0.33 & 0.086 & 0.21 & 
0.066 &  0.031 \\
Random & 0.44 & 0.32 & 0.094 & 0.24 & 
0.077 &  0.028 \\
S$_{(001)}$1 & 0.51 & 0.35 & 0.091 & 0.23 & 
0.079 &  0.021 \\
\end{tabular}
\end{table}

\end{document}